\documentclass[aps,prl,superscriptaddress, twocolumn]{revtex4-1}
\usepackage{lineno,hyperref}
\usepackage{epstopdf}
\usepackage{epsfig}
\usepackage{epstopdf}
\usepackage{graphicx}
\usepackage{bmpsize}
\usepackage{amsfonts}
\usepackage{multirow}

\newcommand{\ben}{\begin{eqnarray}}
\newcommand{\een}{\end{eqnarray}}

\newcommand{\bef}{\begin{figure}[h!bt]\centering}
\newcommand{\eef}{\end{figure}}
\newcommand{\eet}{\end{table}}

\begin{document}
%%%%%%%%%%%%%%%%%%%%%%%%%%%%%%%

\title{Fermi surface topology and negative longitudinal magnetoresistance observed in NbAs$_2$ semimetal}
\author{Bing Shen}
\affiliation{Department of Physics and Astronomy and California NanoSystems Institute, University of California, Los Angeles,
CA 90095, USA}
\author{Xiaoyu Deng}
\affiliation{Department of Physics and Astronomy, Rutgers University, Piscataway, NJ 08854, USA}
\author{Gabriel Kotliar}
\affiliation{Department of Physics and Astronomy, Rutgers University, Piscataway, NJ 08854, USA}
\author{Ni Ni}
\email{Corresponding author: nini@physics.ucla.edu}
\affiliation{Department of Physics and Astronomy and California NanoSystems Institute, University of California, Los Angeles, CA 90095, USA}
\begin{abstract}
We report transverse and longitudinal magneto-transport properties of NbAs$_2$ single crystals. Attributing to the electron-hole compensation, non-saturating large transverse magnetoresistance reaches up to 8000 at 9 T at 1.8 K with mobility around 1 to 2 $\rm m^2V^{-1}S^{-1}$. We present a thorough study of angular-dependent Shubnikov-de Haas (SdH) quantum oscillations of NbAs$_2$. Three distinct oscillation frequencies are identified. First-principles calculations reveal four types of Fermi pockets: electron $\alpha$ pocket, hole $\beta$ pocket, hole $\gamma$ pocket and small electron $\delta$ pocket. Although the angular dependence of $\alpha$, $\beta$ and $\delta$ agree well with the SdH data, it is unclear why the $\gamma$ pocket is missing in SdH. Negative longitudinal magnetoresistance is observed which may be linked to novel topological states in this material although systematic study is necessary to ascertain its origin.

\end{abstract}
\date{\today}
\maketitle
%%%%%%%%%%%%%%%%%%%%%%%%%%%%%%%
Materials with nontrivial topology in their electronic structure often display unusual magneto-transport behavior. Recently large, linear, unsaturating transverse magnetoresistance (TMR) has appeared in Dirac semimetals Cd$_3$As$_2$, Na$_3$Bi and Weyl semimetal TaAs family \cite{TianLiang,JXiong,XCHuang,CZhang}. Negative longitudinal magnetoresistance (NLMR) has been discovered in Na$_3$Bi, TaAs and Cd$_3$As$_2$ \cite{JXiong, XCHuang,CZhang, cd3as2}. In these semimetals, electronic structures exhibit accidental band crossings protected by symmetry and linear energy-momentum dispersion near the Fermi level is observed. Due to their non-trivial topological state, exotic phenomena, such as Fermi-arc surface states, negative longitudinal magnetoresistivity (NLMR) have been observed \cite{ZWang, Weng, Su, lv, JXiong}. Since then, nonmagnetic semimetals with extremely large TMR have re-inspired a lot of research interest because they provide a candidate pool to search for new quantum phases arising from nontrivial topology. NbSb$_2$ is one of the materials showing TMR up to 1300 at 1.8 K under 9 T. Dirac points were suspected in this material \cite{K F Wang, PAband}. However, no further study has been made to understand its Fermi surface topology and examine whether phenomena caused by non-trivial topology exist. In this paper, we study the magneto-transport behavior of NbAs$_2$ single crystals. Large TMR up to 8000 appears, the Fermi surface topology and NLMR are revealed.
\begin{figure}
  \centering
  \includegraphics[width=3.5in]{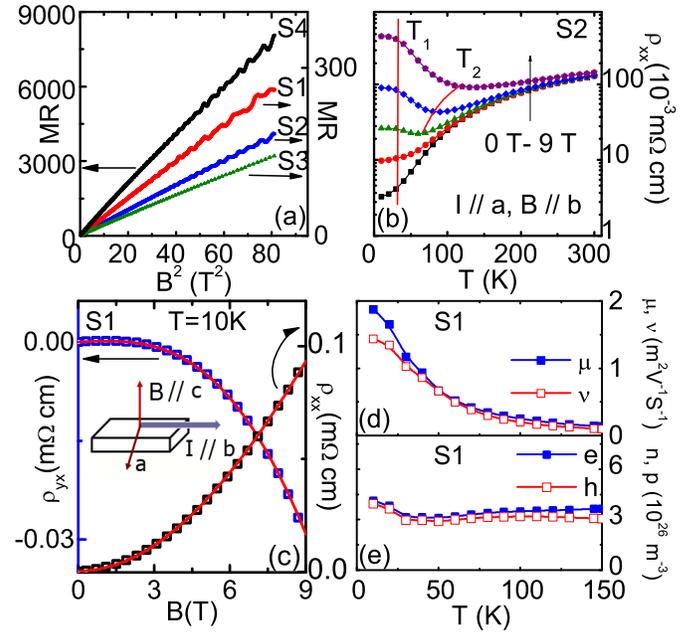}
  \caption{(a) Transverse magnetoresistance vs. $B^2$ at 1.8 K for sample S1, S2, S3 and S4. (b) Temperature dependent transverse resistivity $\rho_{xx}$ of sample S2 measured at 0 T, 1 T, 2 T, 4 T and 9 T with $I // a$ and $B // b$. $T_1$ is the temperature where the resistivity flattening occurs and $T_2$ is the temperature where the minima of $\rho_{xx}$ appears. (c) Field dependent $\rho_{xx}$ and Hall resistivity $\rho_{yx}$ taken at 10 K on S1. The red solid lines are the two-band model fitting curves. Inset: measurement geometry. (d)-(e) Temperature dependent mobility and carrier density of S1, respectively.}
  \label{fig:Fig1}
\end{figure}

NbAs$_2$ single crystals were grown via chemical vapor transport method \cite {supp}. Both X-ray diffraction and wave-length dispersive spectroscopy were used to confirm the phase. Electrical transport measurements were performed using Quantum Design Physical Properties Measurement System (QD PPMS Dynacool). In all measurements, we used x-ray diffraction to determine the crystal orientation first (Fig. S1 in the supplementary material (SM)\cite {supp}). We then shaped the samples into thin rectangular bars and adopted the standard 6-probe configuration for electrical measurements.

First-principles calculations based on density functional theory (DFT) were carried out to study the electronic structure of NbAs$_2$. The full-potential linearized augmented plane-wave method and the generalized gradient approximation of the exchange-correlation potential as implemented in Wien2k package were used \cite{wien2k, PBE}. the generalized gradient approximation of the exchange-correlation potential as implemented in Wien2k package were used \cite{wien2k, PBE}.
Spin-orbit coupling was included in all calculations. The crystallographic structure was taken from Ref. \cite{nbas2}, which is described by the centrosymmetric monoclinic C 1 2/m 1 space group with $a=9.368 \AA, b=33.96\AA, c=7.799 \AA$, and the monoclinic angle between $a$ axis and $c$ axis is $\beta=119.42^{\circ}$. The DFT calculations were performed on a primitive cell with two formula of NbAs$_2$, as well as a conventional cell with four formula of NbAs$_2$ \cite{supp}.

Figure 1(a) shows the field dependent TMR, which is defined as $MR(B) = [\rho_{xx}(B)-\rho_{xx}(0)]/\rho_{xx}(0)$. TMR shows roughly a $B^2$ dependence for all samples independent of the current direction. At 1.8 K under 9 T, TMR reaches 230 for S1, $170$ for S2, 143 for S3 and 8000 for S4. Figure 1(b) shows the temperature dependent transverse resistivity $\rho_{xx}$ of NbAs$_2$ with the current along the $a$ axis ($I//a$) and the field along the $b$ axis ($B//b$) for sample S2. Upon decreasing temperature, the zero field $\rho_{xx}$ decreases with a residual resistivity in the $\mu \Omega \rm cm$ range. As shown in Fig. 1(b), above 2 T, with cooling, $\rho_{xx}$ decreases first, then increases and finally saturates at low temperature, resulting in a resistivity minima at $T_2$ and a flattening below $T_1$. With elevated $B$, $T_2$ moves to higher temperature while $T_1$ remains almost the same. This field induced upturn of resistivity (so-called transformative turn-on temperature behavior) has been observed in various semimetals with extremely large TMR, such as TaAs, WTe$_2$, and its origin is under debate \cite{Gr, PdCoO, MNAli, IPletikosi,YLWang}.

To understand the mechanism of the extremely large TMR data in NbAs$_2$, we performed field dependent transverse magnetoresistivity ($\rho_{xx}$) and Hall resistivity ($\rho_{yx}$) measurements at various temperatures for S1 with $I//b$. Figure 1 (c) shows the representative $\rho_{xx}$ and $\rho_{yx}$ data of S1 taken at 10 K. The nonlinear Hall resistivity observed at 10 K indicates multiband effect in the system. A semiclassical two-band isotropic model is used to analyze the data \cite{MNAli}. We simultaneously fit both $\rho_{xx}$ and $\rho_{yx}$ data using $n, p, \mu$ and $\nu$ as variables, where $n(p)$ and $u(v)$ are the carrier density and mobility of electrons (holes), respectively \cite {supp}. The red solid lines in Fig. 1 (c) are the fitting curves, showing a very good agreement. Figure 1(d) shows the resulting temperature dependent $\mu, v$ and $n, p$ for S1. With decreasing temperature, mobility $\mu$ and $\nu$ increase drastically and show similar strong temperature dependence. The magnitudes of $\mu$ and $\nu$ are comparable to each other for the temperatures from 150 K to 10 K with the largest value of 1$\sim$2 $\rm m^2~V^{-1}S^{-1}$ at 10 K. Charge carrier densities $n$ and $p$ are also close to each other closely, but contrary to the strong temperature dependence in $\mu$ and $\nu$, they are almost temperature independent and the magnitude of them is in the 10$^{26}$ m$^{-3}$ range. Thus electrons and holes are well compensated in NbAs$_2$, which could be responsible for the extremely large TMR. The temperature dependence of mobilities and charge carrier densities resembles the ones in the prototypical semimetal, Bi which also shows extremely large TMR \cite{Bi1, Bi2}.

 \begin{figure}
  \centering
  \includegraphics[width=3.5in]{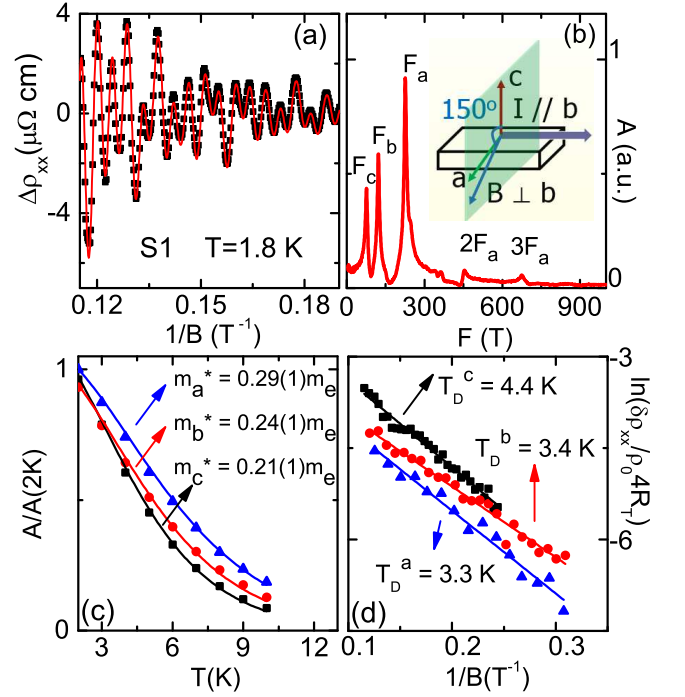}
  \caption{(a) $\Delta\rho_{xx}$, the total oscillation pattern after a polynomial background subtraction, vs. $1/B$ measured at 1.8 K with $1/B$ up to 0.19 (T$^{-1}$). Experimental data (dots); Reconstructed curve (line). The oscillations observed at 0.19$\leq 1/B \leq$0.30 is shown in Fig. S3(a). The measurement geometry is depicted in the inset of (b). $B$ is 150$^\circ$ away from the $c$ axis in the $ac$ plane. (b) The FFT spectrum of $\Delta\rho_{xx}$ at 1.8 K. Inset: The measurement geometry. (c) The normalized temperature dependent amplitude of the respective oscillation, $\delta\rho_{xx}$, associated with $F_a$, $F_b$ and $F_c$. Solid line: fitting. (d) The Dingle plots of the respective $\delta\rho_{xx}$ associated with $F_a$, $F_b$ and $F_c$. Solid line: fitting. $\rho_0$ is the residual resistivity at 0 T and $R_T=\frac{\alpha Tm^*/B}{ sinh(\alpha Tm^*/B)}$.}
  \label{fig:Fig3}
\end{figure}

\begin{figure*}
  \centering
  \includegraphics[width=6.5in]{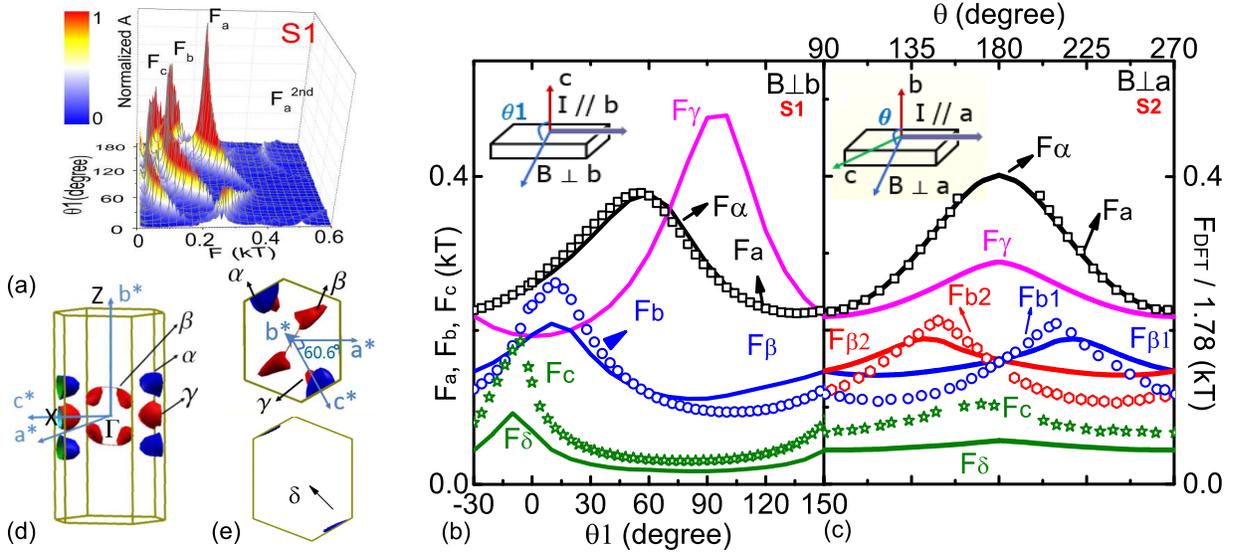}
  \caption{(a) A 3D plot of the FFT spectra of $\Delta\rho_{xx}$ taken at 1.8 K for S1. (b)-(c): The angular dependence of oscillation frequencies. Solid lines are the frequency calculated by DFT, $F_{DFT}$, with a scaling factor 1/1.78. Symbols are the frequencies determined by the SdH measurements. Inset of (b): measurement geometry. $B$ rotates in the $ac$ plane and $\theta 1$ is defined as the rotation angle away from the $c$ axis. Inset of (c): measurement geometry. $B$ rotates around the $a$ axis and $\theta$ is defined as the rotation angle away from the $b$ axis. (d)-(e): The side-view and the top-view of the Fermi surface, respectively \cite{supp}.
 }
  \label{fig:Fig1}
  \end{figure*}

To investigate the Fermi surface topology, we performed angular magneto-transport measurements at low temperatures. As a representative, Figure. 2 presents our analysis on one set of SdH data taken on S1 at 1.8 K with the geometry shown in the inset of Fig. 2(b) \cite{supp}. After subtracting a polynomial background from $\rho_{xx}$, obvious quantum oscillations appear above 3 T in the total oscillation (denoted as $\Delta\rho_{xx}$). Figure 2(b) presents the Fast Fourier transformation (FFT) spectrum of $\Delta\rho_{xx}$. Three obvious fundamental oscillation frequencies $F_a$, $F_b$ and $F_c$ are identified. The oscillation frequency and the extreme cross section $S_k$ are related by the Onsager relation $F=\hbar S_k/2\pi e$ \cite{Onsager relation}. Therefore, to obtain information for each Fermi pocket, we used frequency filtering and inverse FFT method to extract the respective oscillation pattern associated with each frequency \cite{lilu, supp} (denoted as $\delta\rho_{xx}$). To check the reliability of this extraction, we reconstructed $\Delta\rho_{xx}$ by summing respective $\delta\rho_{xx}$. Figure 2(a) shows good agreement between the reconstructed (solid line) and experimental oscillation (black dot). The amplitude of each $\delta\rho_{xx}$ can be expressed by the Lifshitz-Kosevich (LK) formula as $A(B, T)\propto \frac{\alpha Tm^*/B}{ sinh(\alpha Tm^*/B)} exp(-\alpha T_Dm^*/B)$ \cite{Onsager relation}.
Here $\alpha=2\pi^2k_Bm_e/e\hbar=14.69 ~ \rm{T/K}$, $m^*$ is the cyclotron effective mass, and $T_D$ is the Dingle temperature which is related to the scattering rate $\tau$ by $T_D=\frac{\hbar}{2\pi k_B\tau}$. At a fixed $B$, by extracting the amplitude of the respective $\delta\rho_{xx}$ at various temperatures, we obtained Fig. 2(c). The fitting results are $m_a^*=0.29(1)m_e$, $m_b^*=0.24(1)m_e$ and $m_c^*=0.21(1)m_e$. At a fixed $T$, by extracting respective $\delta\rho_{xx}$ at various $B$ for each frequency, we made Fig. 2(d). The obtained $T_D$ from fitting is $T_D^a=3.3~\rm K$, $T_D^b=3.4~\rm K$ and $T_D^c=4.4~\rm K$.

 To map out fine structures of the Fermi surface, we rotated $B$ in the $ac$ plane of S1 around its $b$ axis. The rotation geometry is depicted in the inset of Fig. 3 (b), with $\theta1$ the rotation angle away from the $c$ axis. Figure 3(a) shows the 3D map of the FFT spectra for $\Delta \rho_{xx}$ measured at 1.8 K. Strong angular dependence of the oscillation frequencies $F_{SdH}$ is observed and presented in Fig. 3(b). While two other works on NbAs$_2$ revealed either one or two frequencies with the effective mass ranging from $0.2~m_e$ to 0.37 $m_e$ \cite{tanbas2, nbas2jia}, three distinct frequencies $F_a$, $F_b$ and $F_c$ are identified based on Fig. 3(b). Further information can be extracted from Fig. 3 (c). It describes the angular-dependent frequency of S2 with $B$ rotating around the $a$ axis, where $\theta$ is the rotation angle away from the $b$ axis. Four distinct fundamental frequencies, $F_a$, $F_{b1}$, $F_{b2}$ and $F_c$, appear in Fig. 3 (c). The clear correlation between $F_{b1}$ and $F_{b2}$ (Fig. 3(c)) suggests that they arise from the same type of Fermi pocket.

 DFT calculations were performed to investigate the Fermi surface topology and compared with the SdH experiments. Calculations using both primitive cell and conventional cell gave consistent results \cite{supp}. We found four types of Fermi surfaces in both calculations, which are shown in the reciprocal conventional cell (Figs. 3(d)-(e)) \cite{supp}: (i) two electron pockets $\alpha$ near X point in nearly perfect elliptical shape, (ii) four anisotropic hole pockets $\beta$ near $\Gamma$ and away from the BZ boundary, iii) one hole pocket $\gamma$ in nearly perfect elliptical shape centered at X point, iv) two small electron pockets $\delta$ centered close to X points, which are difficult to see in Fig. 3(d), but better shown in Fig. 3(e). Based on the rotation geometries shown in the inset of Fig3. 3(b) and (c), we computed the frequency $F_{DFT}$ of each pocket using SKEAF \cite {skeaf}. Since the magnitudes of $F_{DFT}$ are larger than the ones of $F_{SdH}$, for a better comparison, $F_{DFT}$/1.78 is plotted in Fig. 3(b) and (c), where $F_{\beta1}$ and $F_{\beta2}$ originate from two different pairs of $\beta$ pockets. As a sanity check, at both $\theta1$=150.6$^{\circ}$ (Fig. 3(b)) and $\theta$=90$^{\circ}$ (Fig. 3(c)), B is perpendicular to the $a^*b^*$ plane (Fig. 3(e)), therefore, $F_{SdH}$ at these two angles should equal, and so should $F_{DFT}$. This is indeed the case as shown in Figs. 3(b) and (c).  Table I summarizes the oscillation frequency and effective mass of several special directions obtained from SdH and DFT.
 \begin{table}[!htbp]
 \caption{\label{tab:test}The comparison of electronic structure parameters of the
Fermi pockets in NbAs$_2$ obtained from experiment and DFT calculations. $F$ is in kT. *1 means in the $a^*b^*$plane and *2 means in the $a^*c^*$ plane.}
 \begin{tabular}{|c|c|c|c|c|c|c|c|}
 \hline
 Fermi pocket &  $F_a$   & $F_b$ & $F_c$ & $F_{\alpha}$ & $F_{\beta}$ & $F_{\gamma}$& $F_{\delta}$ \\
\hline
   Frequency$^{*1}$& 0.226 &0.122 & 0.076 & 0.401& 0.261&0.389 &0.076 \\
   \hline
   $m^*/m_e^{*1}$& 0.29  &0.24  & 0.21 &0.29 & 0.64&0.47 &0.25\\
   \hline

   Frequency$^{*2}$& -- &0.159 & 0.102&0.402 &0.102 &0.290&0.057 \\
   \hline
   $m^*/m_e^{*2}$& -- & 0.3  & 0.26 &0.45 &0.76 &0.71 &0.29\\
   \hline

\end{tabular}
\end{table}

 We notice that in both Figs. 3(b) and (c), the angular dependence of $F_a$ matches $F_{\alpha}$ well. The maxima of $F_a$ and $F_{\alpha}$ is at $\theta1 \sim 60^{\circ}$ where $B//a^*$. This is consistent with the fact that $\alpha$ pocket elongates along the normal of the $a^*b^*$ plane (Fig. 3(e)). Furthermore, just like $F_{b1}$ and $F_{b2}$, at $\theta \sim 90^{\circ}$, $F_{\beta1}$ intersects with $F_{\beta2}$. Therefore, we assign $F_{b1}$ and $F_{b2}$ to the hole pocket $\beta$. As a result, $F_c$ has to be assigned either to the hole $\gamma$ pocket or to the small electron $\delta$ pocket. Since the angular dependence and the size of $F_c$ and $F_{\delta}$ are similar, we tentatively assign $F_c$ to the electron $\delta$ pocket. We are aware that a DFT work suggests that our $F_c$ frequency may come from the $\delta$ pocket \cite {cao}.

 For both rotations, the angular dependencies of $F_{SdH}$ and $F_{DFT}$ of the $\alpha$, $\beta$ and $\delta$ pockets agree well, although the absolute values of them are off. The discrepancy is conceivable since the size of all pockets is small.
 However, it is unclear why the $\gamma$ pocket is missing in SdH. $\gamma$ pocket is predicted to have similar size and effective mass as the $\alpha$ pocket. It is thus surprising that we did not detect the corresponding frequency of $\gamma$ even if we have observed the 3$F_a$ oscillation (Fig. 2(a)). Angular resolved photoemission spectroscopy measurement may shed light on this discrepancy.

\begin{figure}
  \centering
  \includegraphics[width=3.2in]{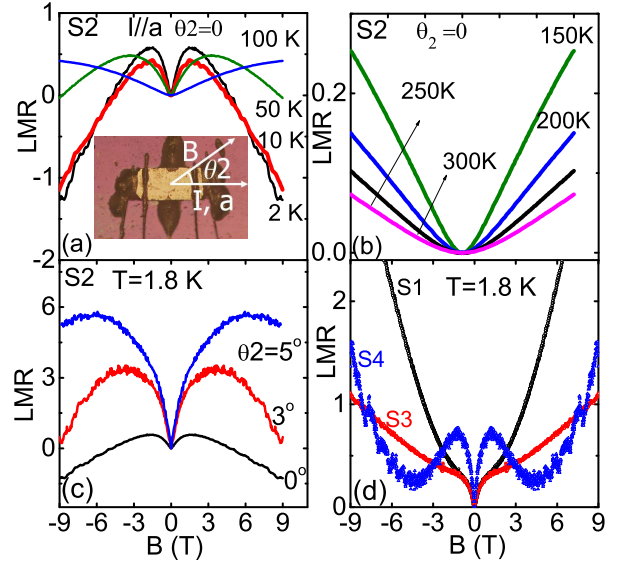}
   \caption{(a) Field dependent LMR taken on S2 at 1.8 K, 10 K, 50 K and 100 K with $B//I//a$. Inset: measurement geometry. $B$ rotates in teh sample plan and $\theta2$ is defined as the rotation angle away from the $a$ axis. (b) Field dependent LMR taken on S2 at 150 K, 200 K, 250 K and 300 K with $B//I//a$. (c) Field dependent LMR taken at 1.8 K at different $\theta2$. (d) Field dependent LMR of S1, S3 and S4 at 1.8 K at $B//I$.}

  \label{fig:Fig1}
\end{figure}

Another novel feature we observed is the NLMR. Figure 4 summarizes the measurements. NLMR clearly appears in S2 with $B$ rotating in the sample plane and $I // a$, where the $\theta2$ is defined as the rotation angle away from the $a$ axis. When $I//a//B$, at 1.8 K, with increasing $B$, LMR first increases up to 0.5 at 1.5 T, then decreases down to its minimum value of -1 at 1.8 K at 9 T (Fig. 4(a)), resulting in a LMR maximum at 1.5 T. This trend of LMR persists up to much higher temperatures with the LMR maximum moving to higher $B$. Up to 9 T, the presence of a MR maxima is still clear at 50 K but is much broadened at 100 K. The overall data pattern suggests a competition between two origins, one with positive and the other with negative response to larger $B$. With even higher temperatures above 150 K, linear LMR is observed up to 9 T (Fig. 4(b)), which may be a consequence of both responses. This trend of NLMR is robust and persists even when the angle between $\theta2$ is 5$^{\circ}$, though with a much weaker negative response (Fig. 4(c)). Figure 4(d) indicates the negative response of S2 is much stronger than the ones in S1, S3 and S4 where S4 has the largest TMR up to 8000 as shown in the inset of Fig. 1(a). Various factors can lead to NLMR \cite{mmbook}. Artifact NLMR can be seen due to asymmetric current flow if the sample size is comparable to the mean free path, poor sample/contact geometry, or the ``current jetting" effect due to the large anisotropy of the material \cite{japan1, japan2}. We have carefully prepared samples to best avoid these effects. S2 is polished into 0.73 mm long, 0.46 mm wide and 0.17 mm thick bar (inset of Fig. 4(a)). The current leads cover the whole area of both edges. The axial anomaly in quasi-two dimensional materials proposed for the NLMR in PtCoO$_2$ is mostly unlikely to be the origin of NLMR here since our sample is quite isotropic suggested by both SdH and DFT data (Table I) \cite{ptcoo2}. Magnetism can cause NLMR, however, no sign of loss of spin scattering appears in our resistivity data and thus impossible to induce such large NLMR. Furthermore, recently it has been proposed that NLMR may occur due to impurity scattering, if the material is in its ultraquantum limit regardless of the band structure \cite{AABurkov,Pallab}. However, the negative response clearly shows even at 2 T (Fig. 4(a)), which is far from the ultraquantum limit here. NLMR can also arise from the chiral anomaly if Weyl nodes are created under external field, which is a charge pumping effect between different Weyl branches \cite{JXiong, CZhang, cd3as2,chiral,nbasjia}. However, careful examination of the band structure and symmetry characterization under field are needed to support this hypothesis. What's more, although great effort has been made to avoid the artifact effect, a systematic study of LMR on samples with different thickness down to tenths of $\mu m$ size is urged to understand the NLMR here \cite{japan2}.

In conclusion, NbAs$_2$ is a compensated semimetal with large mobilities, leading to the observed large MR. Three Fermi pockets are identified and their Fermi topology are mapped out both SdH measurements and DFT calculations. Although the oscillations associated with the hole $\gamma$ pocket are missing, our DFT calculations are overall consistent with the SdH experiment. NLMR exists and further systematic investigation is needed to discern the origin of the observed NLMR.

Note: During the submission of this paper, we noticed serval magneto-transport work on TaSb$_2$, TaAs$_2$ \cite{taas2a,tanbas2,tasb2,nbas2jia,YKLuoN} and TMR data on NbAs$_2$ \cite{tanbas2, nbas2jia}.

\section{Acknowledgments}
Work at UCLA was supported by the U.S. Department of Energy (DOE), Office of Science, Office of Basic Energy Sciences under Award Number DE-SC0011978. Work at Rutgers was supported by the NSF DMREF program under the award NSF DMREF project DMR-1435918. Ni Ni thanks the useful discussion with B. A. Bernevig and J. Xiong.

\end{document}